\documentclass[prc,preprintnumbers,onecolumn]{revtex4}
\usepackage{graphicx}
\usepackage{amssymb}
\usepackage{amsmath}

\begin{document}

\title{Derivation of reaction cross sections from experimental elastic
backscattering probabilities}
\author{V.V.Sargsyan$^{1,2}$, G.G.Adamian$^{1}$, N.V.Antonenko$^{1}$, and
P.R.S.Gomes$^3$}
\affiliation{$^{1}$Joint Institute for Nuclear Research, 141980 Dubna, Russia\\
$^{2}$International Center for Advanced Studies, Yerevan State University,
0025 Yerevan, Armenia\\
$^{3}$Instituto de Fisica, Universidade Federal Fluminense, Av. Litor\^anea,
s/n, Niter\'oi, R.J. 24210-340, Brazil}
\date{\today }

\begin{abstract}
The relationship between the backward elastic scattering probabilities and
reaction cross sections is derived. This is a a very simple and useful
method to extract reaction cross sections for heavy ion systems. We compare
the results of our method with those using the traditional full elastic
scattering angular distributions, for several systems, at energies near and
above the Coulomb barrier. From the calculated reaction and capture cross
sections using the present method, we derive the cross sections of other
mechanisms for  nearly spherical systems.
\end{abstract}

{\pacs{25.70.Jj, 24.10.-i, 24.60.-k  \\
Key words:
reaction
cross
section,
elastic
scattering
probability
at
backward
angle,
quasielastic
scattering
excitation
function
at
backward
angle,
capture
cross section%
}
\maketitle

{}

\section{Introduction}
For a long time, measurements of elastic scattering angular distributions
covering full angular ranges and optical model analysis have been used for
the determination of reaction cross sections. The traditional method
consists in deriving the parameters of the complex optical potentials which
fit the experimental elastic scattering angular distributions and then to
derive the reaction cross sections predicted by these potentials. This can
be done because there is well known and clear relationship between the
reaction and the elastic scattering processes due to the conservation of the
total reaction flux. Any loss from the elastic scattering channel directly
contributes to the reaction channel and vice versa. The direct measurement
of the reaction cross section is a very difficult task, since it would
require the measurement of individual cross sections of all reaction
channels, and most of them could be reached only by specific experiments.
This would require different experimental setups not always available at the
same laboratory and, consequently, such direct measurements would demand a
large amount of beam time and would take probability some years to be
reached. On the other hand, the measurement of elastic scattering angular
distributions is much simpler than that. Even so, both the experimental part
and the analysis of this latter method are not so simple. In the present
work, as an extension of previous works of our group \cite%
{Sargsyan13a,Sargsyan13b}, we present a much simpler method to determine
reaction cross sections than the one using full elastic scattering angular
distribution data. It consists of measuring only elastic scattering at one
backward angle and from that the extraction of the reaction cross sections
can be easily done.

The paper is organized in the following way. In Sec.~II we derive the
formula for the extraction of the reaction cross sections by employing the
experimental elastic scattering excitation function at backward angle. In
Sec.~III we use this formula to extract the reaction cross sections for
several systems and then we compare the results with those extracted from
the experimental elastic scattering angular distributions for the same
systems ($^{4}$He + $^{92}$Mo, $^{4}$He + $^{110,116}$Cd, $^{4}$He + $%
^{112,120}$Sn, $^{6,7}$Li + $^{64}$Zn, and $^{16}$O + $^{208}$Pb). In this
section we also show the comparison of calculated and experimental capture
cross section for the $^{6,7}$Li + $^{64}$Zn systems, and we predict the
approximate cross sections for transfer +\ inelastic processes for those
systems. In Sec.~IV the paper is summarized.

\bigskip

\section{Relationship between reaction cross sections and elastic scattering
excitation function at backward angle}
Quasi-elastic scattering is defined as the sum of elastic scattering,
inelastic excitations and a few nucleon transfer reactions. So, one defines
the quasi-elastic scattering probability as\bigskip \bigskip\
\begin{equation}
P_{qe}(E_{\mathrm{c.m.}},J)=P_{el}(E_{\mathrm{c.m.}},J)+P_{in}(E_{\mathrm{%
c.m.}},J)+P_{tr}(E_{\mathrm{c.m.}},J),
\end{equation}
\bigskip where $P_{el}$, $P_{in}$, and $P_{tr}$ are the elastic scattering,
inelastic and transfer probabilities, respectively. The total reaction
probability may be written as
\begin{equation}
P_{R}(E_{\mathrm{c.m.}},J)=P_{in}(E_{\mathrm{c.m.}},J)+P_{tr}(E_{\mathrm{c.m.%
}},J)+P_{cap}(E_{\mathrm{c.m.}},J)+P_{BU}(E_{\mathrm{c.m.}},J)+P_{DIC}(E_{%
\mathrm{c.m.}},J),
\end{equation}%
where $P_{R}$ refers to the non-elastic reaction channel probability, $%
P_{cap}$\ is the capture probability
(sum of  evaporation-residue formation, fusion-fission, and
quasi-fission probabilities or sum of  fusion and quasi-fission probabilities),
$P_{DIC}$ is the deep inelastic collision
probability, and $P_{BU}$ is the breakup probability, important particularly
when weakly bound nuclei are involved in the reaction \cite{Canto06}. Note
that the deep inelastic collision process is only important at large
energies above the Coulomb barrier.
The deep inelastic collision process  one can neglect
because we are concerned with low energy region.

From the conservation of the total reaction flux one can write \cite%
{Sargsyan13a,Canto06} the expression
\begin{equation}
P_{el}(E_{\mathrm{c.m.}},J)+P_{R}(E_{\mathrm{c.m.}},J)=1
\end{equation}
or
\begin{equation}
P_{qe}(E_{\mathrm{c.m.}},J)+P_{cap}(E_{\mathrm{c.m.}},J)+P_{BU}(E_{\mathrm{%
c.m.}},J)=1.  \label{cn_eq}
\end{equation}
%
%
Here and in the following of this paper, we neglect the deep inelastic
collision, since we are concerned with low energies.\ Thus, one can extract
the reaction probability $P_{R}(E_{\mathrm{c.m.}},J=0)$ at $J=0$ from the
experimental elastic scattering probability $P_{el}(E_{\mathrm{c.m.}},J=0)$
at $J=0$:
\begin{equation}
P_{R}(E_{\mathrm{c.m.}},J=0)=1-P_{el}(E_{\mathrm{c.m.}},J=0)=1-d\sigma
_{el}(E_{\mathrm{c.m.}})/d\sigma _{Ru}(E_{\mathrm{c.m.}}).  \label{1cn_eq}
\end{equation}
Here, the elastic scattering probability~\cite%
{Canto06,Timmers,Timmers1,Timmers2,Zhang}
\begin{equation}
P_{el}(E_{\mathrm{c.m.}},J=0)=d\sigma _{el}/d\sigma _{Ru}
\end{equation}
for angular momentum $J=0$ is given by the ratio of the elastic scattering
differential cross section and Rutherford differential cross section at  180
degrees. Furthermore, one can approximate the $J$ dependence of the reaction
probability $P_{R}(E_{\mathrm{c.m.}},J)$ at a given energy $E_{\mathrm{c.m.}%
} $ by shifting the energy \cite{Bala}:
\begin{equation}
P_{R}(E_{\mathrm{c.m.}},J)\approx P_{R}(E_{\mathrm{c.m.}}-\frac{\hbar
^{2}\Lambda }{2\mu R_{b}^{2}}-\frac{\hbar ^{4}\Lambda ^{2}}{2\mu ^{3}\omega
_{b}^{2}R_{b}^{6}},J=0),  \label{2cn_eq}
\end{equation}
where $\Lambda =J(J+1)$, $R_{b}=R_{b}(J=0)$ is the position of the Coulomb
barrier at $J=0$, $\mu =m_{0}A_{1}A_{2}/(A_{1}+A_{2})$ is the reduced mass ($%
m_{0}$ is the nucleon mass), and $\omega _{b}$ is the curvature of the
s-wave potential barrier. Employing Eqs. (\ref{1cn_eq}) and (\ref{2cn_eq}),
converting the sum over the partial waves $J$ into an integral, and
expressing $J$ by the variable $E=E_{\mathrm{c.m.}}-\frac{\hbar ^{2}\Lambda
}{2\mu R_{b}^{2}}$, we obtain the following simple expression:
\begin{equation}
\sigma _{R}(E_{\mathrm{c.m.}})=\frac{\pi R_{b}^{2}}{E_{\mathrm{c.m.}}}%
\int_{0}^{E_{\mathrm{c.m.}}}dE[1-d\sigma _{el}(E)/d\sigma _{Ru}(E)][1-\frac{%
4(E_{\mathrm{c.m.}}-E)}{\mu \omega _{b}^{2}R_{b}^{2}}].  \label{3cn_eq}
\end{equation}%
%
%
%
%
%
%
%
%
The formula (\ref{3cn_eq}) relates the reaction cross section with elastic
scattering excitation function at backward angle. By using the experimental
elastic scattering probabilities $P_{el}(E_{\mathrm{c.m.}},J=0)$ and Eq. (%
\ref{3cn_eq}) one can obtain the reaction cross sections.

It is important to mention that since the generalized form of the optical
theorem connects the reaction cross section and forward elastic scattering
amplitude\cite{Canto06}, from our method we show that the forward and
backward elastic scattering amplitudes are related with each other.

\section{Results of calculations}
\subsection{ Reaction cross sections}
In the following, we show the results of our method to extract the reaction
cross section, using Eq. (\ref{3cn_eq}).  To calculate the position $R_{b}$
and frequency $\omega _{b}$ of the Coulomb barrier, we use the
nucleus-nucleus interaction potential $V(R,J)$ of Ref. \cite{Pot}. For the
nuclear part of the nucleus-nucleus potential, the double-folding formalism
with the Skyrme-type density-dependent effective nucleon-nucleon interaction
is employed~\cite{Pot}.

To confirm the validity of our method of the extraction of $\sigma _{R}$,
firstly we compare the obtained reaction cross sections with those extracted
from the traditional experimental elastic scattering angular distributions
plus optical potential method. The results from our method are shown as
solid (red color on-line) and dashed (blue color on-line) lines in all
figures from Fig. 1 to Fig. 6, whereas the results obtained from the
traditional full elastic scattering angular distribution data are shown by
solid squares. As the backscattering elastic data were not taken at 180
degree, but rather at backward angles in the range from 150 to 170 degrees,
the corresponding center of mass energies were corrected by the centrifugal
potential at the experimental angle, as suggested by Timmers et al. \cite%
{Timmers}. In figures 7 and 8 we also show results of our calculations for
the capture cross sections, and other curves are shown.
\begin{figure}[tbp]
\includegraphics[scale=1]{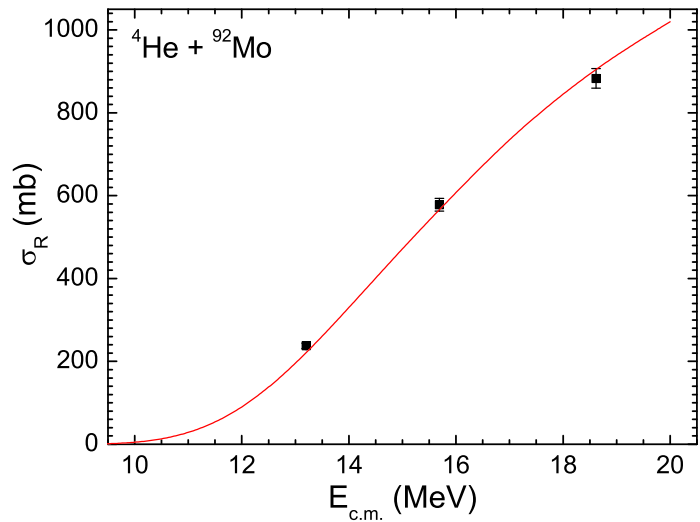}
\caption{(Color on line) The extracted reaction cross sections employing Eq. (\protect\ref%
{3cn_eq}) (solid line) for the $^{4}$He + $^{92}$Mo reaction. The used
experimental elastic scattering probabilities at backward angle are from
Ref. \protect\cite{hemo}. The reaction cross sections extracted from the
experimental elastic scattering angular distribution with optical potential
are presented by   squares \protect\cite{hemo}.
}
\label{1_fig}
\end{figure}
\begin{figure}[tbp]
\includegraphics[scale=1]{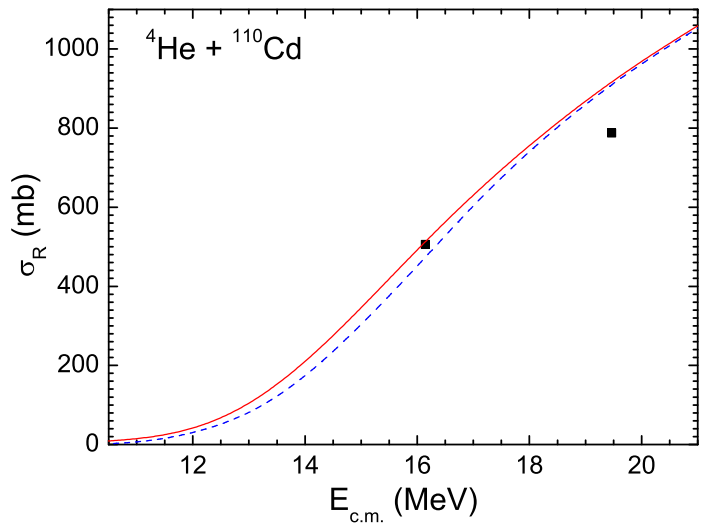}
\caption{(Color on line)  The extracted reaction cross sections employing Eq. (\protect\ref%
{3cn_eq}) (lines) for the $^{4}$He + $^{110}$Cd reaction. The used
experimental elastic scattering probabilities at backward angle are from
Refs. \protect\cite{hecd1,hecd3} (solid line) and Ref. \protect\cite{hecd2}
(dashed lines). The reaction cross sections extracted from the experimental
elastic scattering angular distribution with optical potential are presented
by   squares \protect\cite{hemo}. }
\label{2_fig}
\end{figure}
\begin{figure}[tbp]
\includegraphics[scale=1]{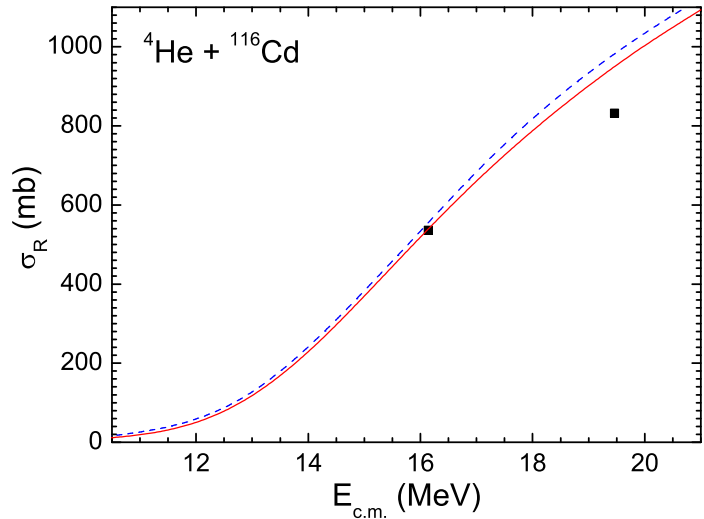}
\caption{(Color on line)  The same as in Fig.~2 but for the $^{4}$He + $^{116}$Cd reaction. }
\label{3_fig}
\end{figure}
\begin{figure}[tbp]
\includegraphics[scale=1]{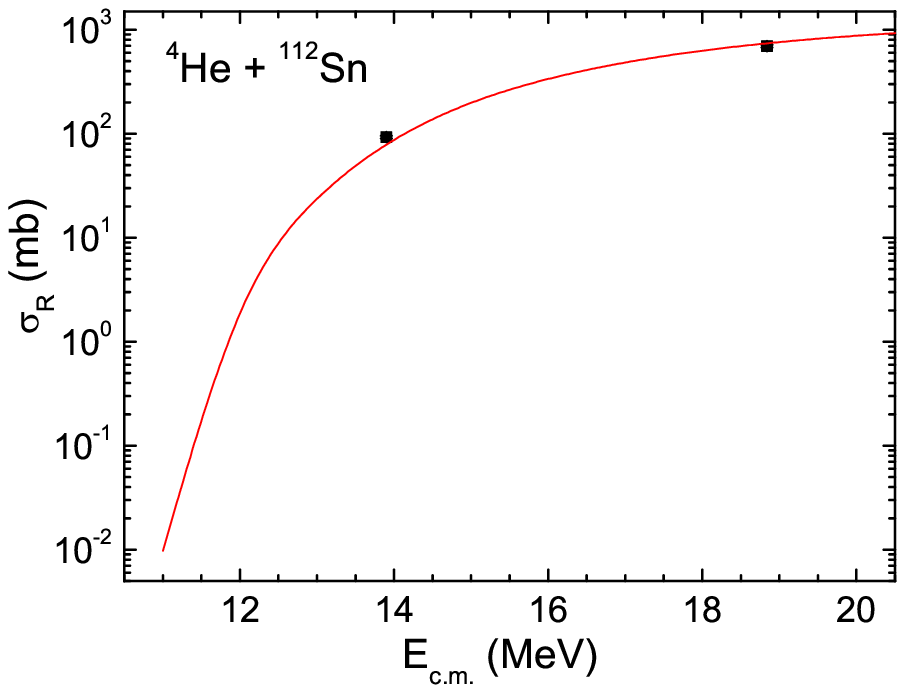}
\caption{(Color on line) The extracted reaction cross sections employing Eq. (\protect\ref%
{3cn_eq}) (solid line) for the $^{4}$He + $^{112}$Sn reaction. The used
experimental elastic scattering probabilities at backward angle are from
Ref. \protect\cite{hemo}. The reaction cross sections extracted from the
experimental elastic scattering angular distribution with optical potential
are presented by  squares \protect\cite{hemo}. }
\label{4_fig}
\end{figure}
\begin{figure}[tbp]
\includegraphics[scale=1]{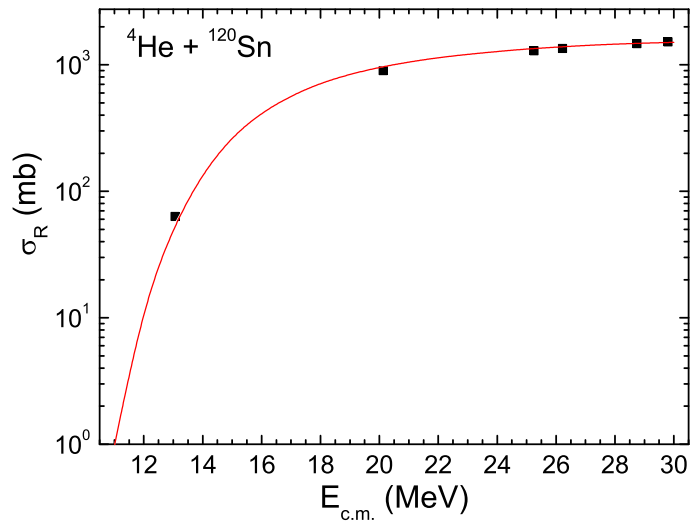}
\caption{(Color on line)  The same as in Fig.~4 but for the $^{4}$He + $^{120}$Sn reaction. }
\label{5_fig}
\end{figure}
\begin{figure}[tbp]
\includegraphics[scale=1]{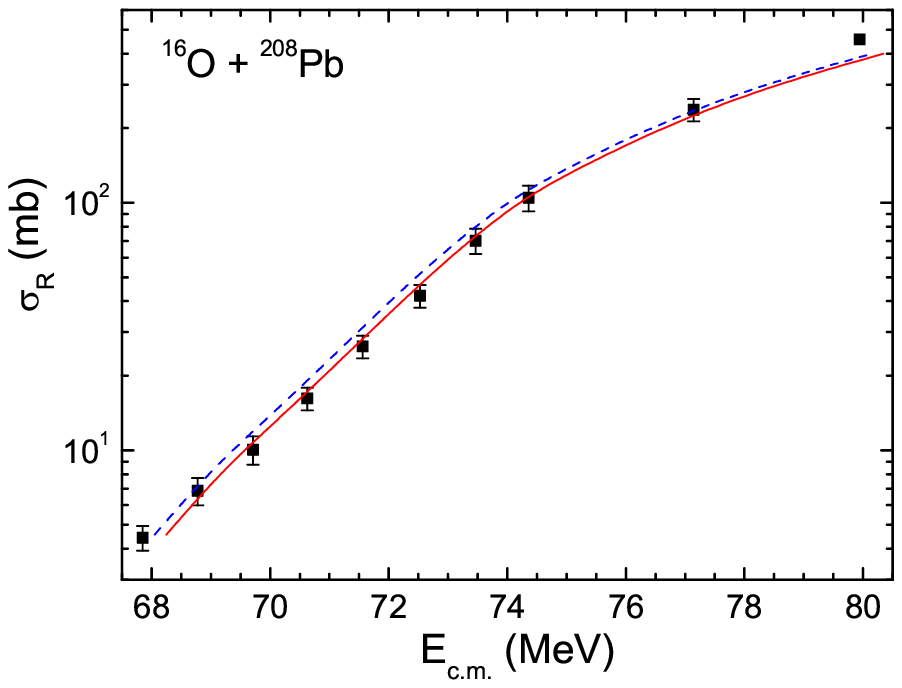}
\caption{(Color on line)  The extracted reaction cross sections employing Eq. (\protect\ref%
{3cn_eq}) (solid line) for the $^{16}$O + $^{208}$Pb reaction. The used
experimental elastic scattering probabilities at backward angle are from
Ref.~\protect\cite{opb}. The reaction cross sections extracted from the
experimental elastic scattering angular distribution with optical potential
are presented by   squares \protect\cite{opb}.}
\label{6_fig}
\end{figure}
As it can be observed in Figs. 1--8, there is a good agreement between
reaction cross sections extracted from experimental elastic scattering at
backward angle and from the experimental elastic scattering angular
distributions with optical potential for the reactions $^{4}$He + $^{92}$Mo,
$^{4}$He + $^{110,116}$Cd, $^{4}$He + $^{112,120}$Sn, $^{16}$O + $^{208}$Pb,
and $^{6,7}$Li + $^{64}$Zn at energies near and above the Coulomb barrier.
One can see that the used formula (\ref{3cn_eq}) is suitable not only for
almost spherical nuclei, but also for the reactions with slightly deformed
target-nuclei. The deformation effect is effectively contained in the
experimental $P_{el}$. For very deformed nuclei, it is not possible
experimentally to separate elastic events from the low lying inelastic
excitations. In our calculations, to obtain better agreement for the
reactions $^{16}$O+$^{208}$Pb and $^{6}$Li+$^{64}$Zn, the extracted reaction
cross sections were shifted in energy by 0.3 MeV to higher energies and 0.4
MeV to lower energies with respect to the measured experimental data,
respectively. There is no clear physical justification for the energy shift.
The most probable reason might be related with the uncertainty
associated with the elastic scattering data.

\subsection{\protect\bigskip Capture and transfer plus breakup plus
inelastic cross sections}
By using a similar formalism as the one presented in Section II and Eq. (\ref%
{cn_eq}), the capture cross section can be written, if one assumes that $%
P_{BU}=0$, since it is much smaller than $P_{qe}$, as \cite{Sargsyan13b}
\begin{equation}
\sigma _{cap}(E_{\mathrm{c.m.}})=\frac{\pi R_{b}^{2}}{E_{\mathrm{c.m.}}}%
\int_{E_{\mathrm{c.m.}}-\frac{\hbar ^{2}\Lambda _{cr}}{2\mu R_{b}^{2}}}^{E_{%
\mathrm{c.m.}}}dE[1-d\sigma _{qe}(E)/d\sigma _{Ru}(E)][1-\frac{4(E_{\mathrm{%
c.m.}}-E)}{\mu \omega _{b}^{2}R_{b}^{2}}],  \label{4cn_eq}
\end{equation}%
where in $\Lambda _{cr}=J_{cr}(J_{cr}+1)$, $J_{cr}$ is the critical angular
momentum at which potential pocket in the nucleus-nucleus interaction
potential $V(R,J)$ vanishes and capture does not occur. So, the capture
cross sections can be extracted from the experimental quasielastic
scattering probabilities $P_{qe}(E_{\mathrm{c.m.}},J=0)=d\sigma
_{qe}/d\sigma _{Ru}$, as it was already demonstrated in Ref. \cite%
{Sargsyan13b}.
\begin{figure}[tbp]
\caption{(Color on line) The extracted reaction (solid line) and capture
(dashed line) cross sections employing Eqs. (\protect\ref{3cn_eq}) and (%
\protect\ref{4cn_eq}) for the $^{6}$Li + $^{64}$Zn reaction. The used
experimental elastic and quasielastic scattering probabilities at backward
angle are from Ref.~\protect\cite{Torresi,Pietro}. The reaction cross
sections extracted from the experimental elastic scattering angular
distribution with optical potential and capture (fusion) cross sections are
presented by   circles \protect\cite{Torresi,Pietro},  triangles \protect\cite{Gomes034,GomesPLB04}
and   squares
\protect\cite{Gomes034,GomesPLB04},   stars \protect\cite{Torresi,Pietro}, respectively.
}
\label{7_fig}\includegraphics[scale=1]{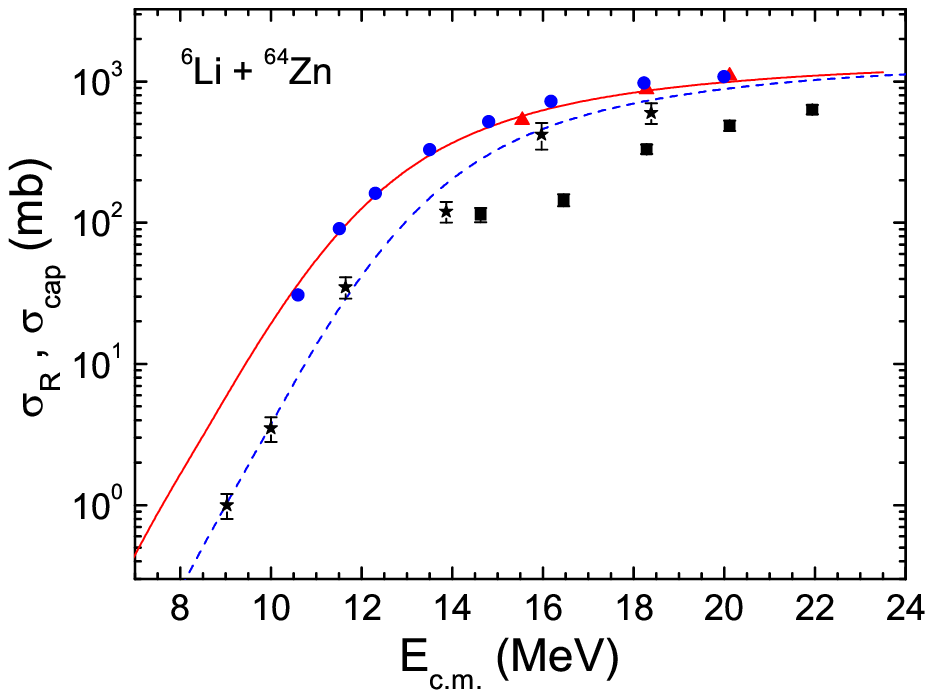}
\end{figure}
\begin{figure}[tbp]
\includegraphics[scale=1]{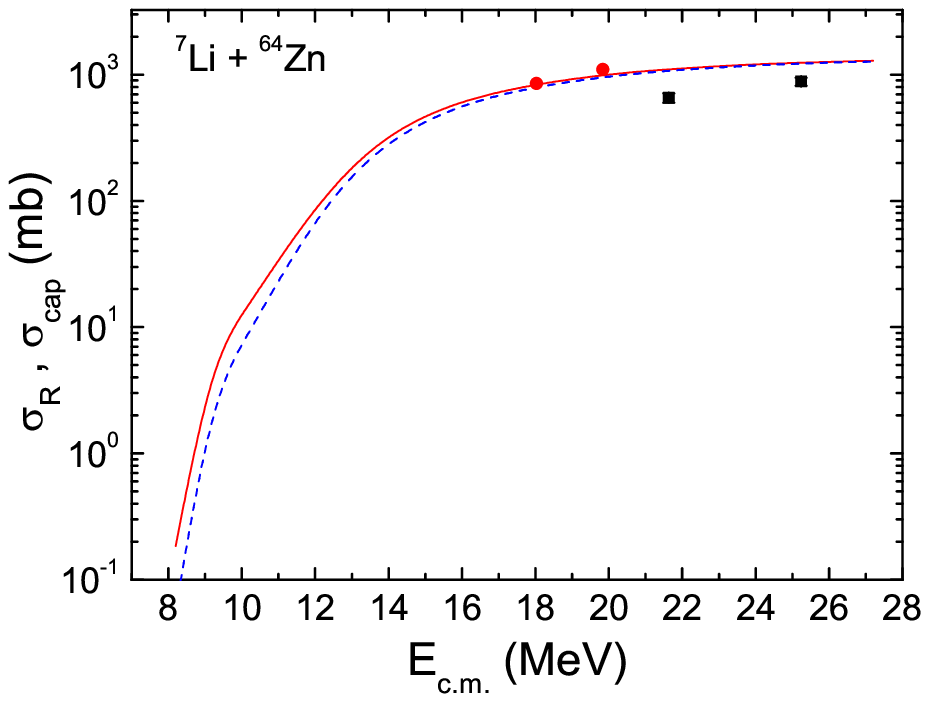}
\caption{(Color on line) The same as in Fig. 7, but for the $^{7}$Li + $%
^{64} $Zn reaction. The reaction cross
sections extracted from the experimental elastic scattering angular
distribution with optical potential and capture (fusion) cross sections are
presented by   circles   \protect\cite{Gomes034,GomesPLB04}
and   squares
\protect\cite{Gomes034,GomesPLB04}, respectively.
}
\label{8_fig}
\end{figure}
In figures 7 and 8 we also show the results of our calculations for capture
cross sections of the $^{6,7}$Li+$^{64}$Zn systems, for which the fusion
process can be considered to exhaust the capture cross section. Figure 7
shows that the extracted and experimental capture cross sections are in good
agreement for the $^{6}$Li+$^{64}$Zn reaction at energies near and above the
Coulomb barrier   for the data taken in Refs. \cite{Torresi,Pietro}.
Note that the extracted capture excitation function is shifted in energy by
0.7 MeV to higher energies with respect to the experimental data.
This could be the result of different energy calibrations in the
experiments on the capture measurement and quasielastic scattering.
The data taken in Refs. \cite{Gomes034,GomesPLB04} are below our
predictions. This fact was already observed and commented in Ref. \cite%
{Gomes09}, and the reason given for the low fusion cross sections was as
owing to experimental problems with the high electronic threshold of the
events, when the data were taken. Figure 8 shows that the capture cross
section for the $^{7}$Li+$^{64}$Zn system, obtained in the same works of
Refs. \cite{Gomes034,GomesPLB04} is also below our predictions. The same
reason for this behavior as for the $^{6}$Li+$^{64}$Zn system was given in
the same Ref. \cite{Gomes09}, since the $^{6}$Li and $^{7}$Li data were
taken at the same experiment.

The extraction  of reaction (capture) cross sections from the
  experimental elastic (quasielastic) backscattering probabilities
  leads to uncertainties of the order of 10\% at energies above
  the Coulomb barrier. At energies below the barrier the uncertainties
  are larger because a deviation of the elastic (quasielastic)
  backscattering cross section from the Rutherford cross section is
  comparable with the experimental uncertainties. Those overall
uncertainties are comparable with the ones obtained from the
traditional method using full elastic scattering angular distributions.

For the $^{7}$Li+$^{64}$Zn reaction, the $Q$-value of the one neutron
stripping transfer is positive and this process should have a reasonable
high probability to occur, whereas for the $^{6}$Li+$^{64}$Zn reaction, $Q$%
-values of neutron transfers are negative. Therefore, one might expect that
transfer cross sections for $^{7}$Li+$^{64}$Zn are larger than for $^{6}$Li+$%
^{64}$Zn. Concerning breakup, since $^{6}$Li has a smaller threshold energy
for breakup than $^{7}$Li, one might expect that breakup cross sections for $%
^{6}$Li+$^{64}$Zn are larger than for $^{7}$Li+$^{64}$Zn. Actually, in Fig.
9 one can observe that our calculations show that $\sigma (^{7}\mathrm{Li}%
+^{64}\mathrm{Zn})>\sigma (^{6}\mathrm{Li}+^{64}\mathrm{Zn})$, where $\sigma
=\sigma _{R}-\sigma _{cap}\approx \sigma _{tr}+\sigma _{in}$ since $\sigma
_{tr}+\sigma _{in}\gg \sigma _{BU}$ for these light systems at energies
close and below the Coulomb barrier ($\sigma_{tr}$,  $\sigma_{in}$, and $\sigma _{BU}$   are
 the transfer,   inelastic  scattering, and breakup cross sections,
 respectively). So, our present method of extracting
reaction and capture cross sections from backward elastic scattering data
allows the approximate determination of the sum of transfer and inelastic
scattering cross sections, or $\sigma _{tr}+\sigma _{in}+\sigma _{BU}$ in
systems where $P_{BU}$ can not be neglected. For both systems investigated,
the values of these cross sections are shown to increase with $E_{\mathrm{%
c.m.}}$, reach a maximum slightly above the Coulomb barrier energy and after
decrease. The difference between the two curves in Fig. 9 may be considered
approximately as the difference of $\sigma _{tr}$\ between the two systems,
since $\sigma _{in}$ should be similar for both systems with the same
target, apart from the excitation of the bound excited state of $^{7}$Li.
Because $\sigma _{tr}(^{7}\mathrm{Li}+^{64}\mathrm{Zn})\gg \sigma _{tr}(^{6}%
\mathrm{Li}+^{64}\mathrm{Zn})$ one can find $\sigma _{tr}(^{7}\mathrm{Li}%
+^{64}\mathrm{Zn})\approx \sigma (^{7}\mathrm{Li}+^{64}\mathrm{Zn})-\sigma
(^{6}\mathrm{Li}+^{64}\mathrm{Zn})$. The maximum absolute value of the
transfer cross section $\sigma _{tr}$ at energies near the Coulomb barrier
is about 30 mb. Fig. 9 also shows that the difference between transfer cross
sections for $^{7}$Li and $^{6}$Li are much more important than the possible
larger  $\sigma _{BU}$ for $^{6}$Li than for $^{7}$Li.

\begin{figure}[tbp]
\caption{The extracted $\protect\sigma_R - \protect\sigma_{cap}$ for the
reactions $^{6}$Li + $^{64}$Zn (dashed line) and $^{7}$Li + $^{64}$Zn (solid
line).}
\label{9_fig}\includegraphics[scale=1]{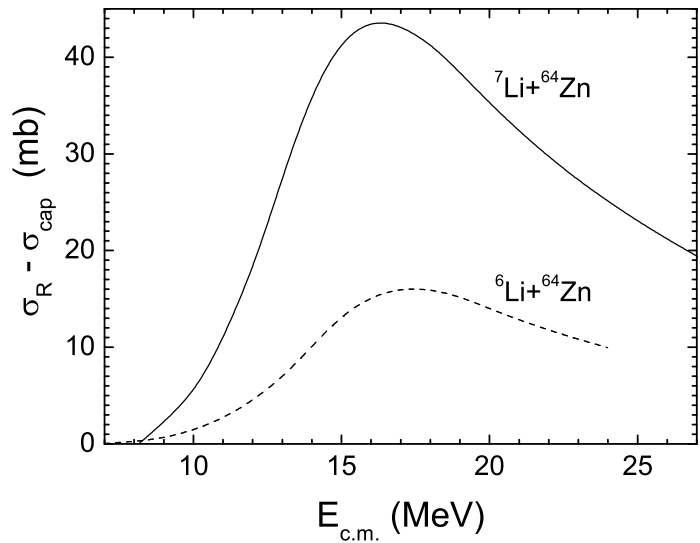}
\end{figure}
%

\section{Summary}
We propose a new and very simple way to determine reaction cross sections,
through a relation (\ref{3cn_eq}) between the elastic scattering excitation
function at backward angle and reaction cross section. We show, for several
systems, that this method works well and that the elastic backscattering
technique could be used as an important and simple tool in the study of the
reaction cross sections. The extraction of reaction (capture)  cross sections from the
elastic (quasielastic)  scattering at backward angle is possible with reasonable
uncertainties as long as the deviation between the elastic (quasielastic) scattering cross
section and the Rutherford cross section exceeds the experimental
uncertainties significantly.
The behavior of the transfer+inelastic
excitation function extracted from the experimental probabilities of the
elastic and quasielastic scatterings at backward angle was also shown.

\vspace*{1cm}

We are grateful to G.~Kiss, R.~Lichtenth\"{a}ler, P.~Mohr,  and M.~Zadro
for providing us the experimental data.
P.R.S.G. acknowledges the partial financial support from CNPq and FAPERJ.
This work was supported by DFG, NSFC, RFBR, and JINR grants.
The IN2P3(France)-JINR(Dubna) and Polish - JINR(Dubna) Cooperation Programmes
are gratefully acknowledged.\newline


\end{document}